# 200 Years of the Navier-Stokes Equation

Sylvio R. Bistafa

Retired, Polytechnic School, University of São Paulo

São Paulo, SP, Brazil

**Abstract**

The year 2022 marked the 200th anniversary of the first appearance of the Navier-Stokes equation, a landmark in Fluid Dynamics introduced by Claude-Louis Navier in 1822. This equation revolutionized the understanding of fluid motion by incorporating viscosity and friction into the equations, expanding their applicability beyond idealized fluids. In this manuscript, we explore the historical development of the Navier-Stokes equation and its profound impact on Fluid Dynamics over the past two centuries. From Navier's initial insights to George Stokes' experimental validations and subsequent contributions by other scientists, we trace the evolution of this equation. We also delve into its practical applications, including its role in the development of Computational Fluid Dynamics. The Navier-Stokes equation has played a pivotal role in advancing our understanding of fluid behavior, making it a cornerstone of modern science and engineering.

1. Introduction

The first derivations of the Navier–Stokes equation appeared in two memoirs by Claude-Louis Navier: *Sur les lois des mouvements des fluides, en ayant égard à l'adhésion des molecules* [1], published in the *Annales de Chimie et de Physique* for the year of 1821 (the printed version actually appeared in 1822), here referred to as the 1$^{st}$ memoir; and *Sur Les Lois du Mouvement des Fluides* [2], which appeared in the *Mémoires de L'Académie Royale des Sciences de L'Institut de France* for the year of 1823 (actually appeared in printing in 1827), here referred to as the 2$^{nd}$ memoir. Nonetheless, according to the records, both were read at *L'Académie* on March 18$^{th}$, 1822, and this was considered the year that marks the appearance of the equation, when it was first publicly announced. These publications formally introduced, for the first time, friction in the equations of fluid motion. Up until then, the equations of motion had been limited to perfect fluids, following the publication of the well-known Euler's equations for non-viscous fluids in 1755.

Navier's inclination for scholarship and his background in higher analysis at the *École Polytechnique* and in practical engineering at the *École des Ponts et Chaussées* put him in the ideal position to make significant contributions to the science of fluid flow by realizing that fluid friction was the main cause for the deviation of experiments from theory. Reference [3] gives a thorough discussion of Navier's development of the Navier-Stokes equation contained in both memoirs. In his developments, Navier begins by adopting Laplace's molecular program that considers bodies in general as made up of particles which are close to each other and which act on each other by means of two opposing forces - one of attraction and one of repulsion - which, when in a state of equilibrium, cancel each other out. When the fluid moves, and all the molecules, being carried away by a common motion, preserve their respective situations, the state of these molecules does not change, and no new action is established in the interior of the fluid. However, when there is difference in velocities between two molecules, the repulsive force between these two molecules will change. According to Navier, a quantity given by the difference in velocities between the two molecules, multiplied by a function of the distance of these two molecules (which decreases very fast as the distance increases), and by a constant relative to the 'adherence of the fluid molecules' (viscosity), gives the repulsive force between the two molecules (force that will be of attraction if this quantity is negative).

After having established the nature of the molecular forces for fluids in motion, Navier makes an analogy between the motion of viscous fluids and the motion of elastic solids, as developed in the 1st memoir, to obtain, for the first time, the incompressible form of the Navier-Stokes equation as it is known today. Because of contradictory experimental results, Navier's main concern thereafter was to know what would be the appropriate boundary conditions to be satisfied at solid boundaries, for cases where the molecules of the walls exert a particular action upon those of the fluid. He then, in the 2nd memoir, by applying Lagrange's method of moments, obtains again the Navier-Stokes equation, but with new boundary conditions.

## 2. What is the Navier–Stokes equation?

The Navier–Stokes equation is a nonlinear partial differential equation, which governs the motion of *real viscous fluids* and can be seen as Newton's second law of motion for fluids. It describes the physics of many flow phenomena of scientific and engineering interest, and it may be used to model the weather, air and ocean currents, pollution dispersion in these fluid media, water flow in tubes, air flow over a wing, blood flow through arteries, etc. It can help engineers to design power stations, aircrafts, cars, fluid flow machines such as pumps, turbines, ventilators, and many other types of equipment.

---

**The Incompressible Navier–Stokes Equation**

$$\underbrace{\rho \left( \frac{\partial \boldsymbol{u}}{\partial t} + (\boldsymbol{u} \cdot \boldsymbol{\nabla}) \boldsymbol{u} \right)}_{\substack{acceleration \\ of\ the\ fluid\ particle \\ of\ velocity\ \boldsymbol{u} \\ (local + convective\ acceleration)}} = \underbrace{\rho \boldsymbol{g}}_{\substack{gravitaional \\ force\ term}} \underbrace{- \nabla p}_{\substack{pressure \\ force\ term}} + \underbrace{\mu \nabla^2 \boldsymbol{u}}_{\substack{viscous \\ force\ term}} ,$$

$\mu$ is viscosity, $\rho$ is density, and quantities in bold are vectors.

The continuity equation $\boldsymbol{\nabla} \cdot \boldsymbol{u} = 0$, together with the three components of the Navier-Stokes equation, form a system of equations necessary to obtain the three velocity components and the pressure.

The boundary condition that accompanies this equation is $\boldsymbol{u} = \boldsymbol{U}$ on solid boundaries, where $\boldsymbol{U}$ is the boundary velocity.

---

The Navier–Stokes equation is an evolution of the Euler's equation. This equation governs the motion of the *perfect non viscous fluid* and as such can be seen as the Navier–Stokes equation without the viscous term $\mu \nabla^2 \boldsymbol{u}$.

At the beginning of the nineteenth century, the knowledge on the phenomena of fluid resistance and flow retardation was empirical. This was also the case of Navier, who even was responsible for the preparation of a revised edition of Belidor's *Architecture hydraulique*, then a very popular practical book among hydraulic engineers. Nonetheless, it was Navier who first tried to insert new terms in Euler's hydrodynamic equations, by applying Laplace's new molecular theory to model viscosity.

In contrast to the empirical approaches then in use, these equations were considered too complex for not offering ready answers to practical problems faced by engineers at the time. Besides, the mathematical contemporary capability was not prepared to tackle these equations. Indeed, they were among the first partial differential equations ever to have been written, and they involved the non-linearity term $(\boldsymbol{u} \cdot \boldsymbol{\nabla})\boldsymbol{u}$ that has troubled solutions to this day. Even if someone had been willing to modify Euler's equations, he would have lacked clues about the structure of the new terms and also because the concept of internal fluid friction was as yet immature [4].

## 3. The first five births of the Navier-Stokes equation

The title of this section was borrowed from Darrigol [4], who in this publication presented a thorough an excellent account on the first proposals of the Navier-Stokes equation, contextualized with the historical developments after and before the appearance of this equation.

Fluid friction forces associated with viscosity were not suitably modelled until well into the nineteenth century. The inclusion of viscous forces into the equations of fluid motion has been first proposed by Claude-Louis Navier in 1822 [1]. The modern theory of elasticity may be considered to have its birth in the same publication, when Navier first gave the equations for the equilibrium and motion of an (isotropic, one-constant) elastic solid. By having proposed what is considered to be the first modern theory of elasticity, Navier soon perceived that these equations could be extended to other continuous media and taking as a starting point the equations for elastic solids, he wrote the equation for the motion of viscous fluids, substituting fluid particle velocities for elastic solid displacements, and the fluid viscosity constant (called 'adherence constant' by Navier) for the elastic solid constant.

Thereafter, more or less independently and by using different arguments, the viscous equations were re-obtained by Augustin Cauchy, Siméon Poisson, and Adhémar Barré de Saint-Venant. As noted by Darrigol [4], each new discoverer either ignored or denigrated his predecessors' contribution. Each had his own way to justify the equation.

As for the involvement of Stokes with the equation of motion for viscous flows is concerned, it began in 1845 when Stokes publishes "On the theories of the internal friction of fluids in motion" [5]. Similar to the approaches of Navier, Stokes used a continuity argument to justify the same equation of motion for elastic solids as for viscous fluids. A practical motivation was that Stokes seems to have realized that the viscosity of the air flowing around the pendulum could play a role in making a pendulum behave differently than the ideal pendulum in a vacuum. By applying methods similar to Cauchy's and Poisson's, Stokes arrived at the Navier-Stokes equation by saying that this equation and the equation of continuity "[…] are applicable to the determination of the motion of water in pipes and canals, to the calculation of friction on the motions of tides and waves, and such questions" [5, p. 93].

Since many investigators had corroborated the equation of motion for viscous flows as developed by Navier, one may wonder why Stokes became also associated with this equation. The answer might be that he made extensive comparisons of theory and experiments of different researchers with cylindrical rods, spheres, spheres at the end of long and short rods, oscillating disks, long and short pendula oscillating in air and water, etc. Therefore, differently from the other authors of the Navier-Stokes equation, and similarly to Navier, Stokes had a very clear intention on the practicality of his efforts by confronting theory with experiments, and this might be a reason why he and Navier became associated with the equation of motion for viscous flows.

## 4. The solutions of the Navier-Stokes equation

As far as the solutions of the Navier-Stokes equation (referred hereafter simply as the N-S equation) is concerned, firstly, it should be pointed out that the N-S equation as posed by their first authors, are strictly only valid for slow motions in capillaries, meaning *laminar flows*. These first derivations did not intend this equation to be used for the most encountered *turbulent flows*. This is a serious restriction because most flows of interest are generally turbulent. Nonetheless, the quest for the solutions of the N-S equation has fostered the development of very creative mathematical methods, which have allowed a deeper understanding of laminar flows, and have sprung the development of numerical solutions for more complex laminar flows and turbulent flows in general.

## 4.1 Exact solutions

The exact solutions of the N-S equation may be divided into two main categories. The first category, include solutions for which the nonlinear term $(\boldsymbol{u} \cdot \boldsymbol{\nabla})\boldsymbol{u} = 0$, owing to the simple nature of the flow. Flows that fall into this category are the Couette flow (from which the lubrication theory is developed), the Poiseuille flow (flow in a tube of uniform cross section), the flow between rotating cylinders, Stokes' first and second problems, and the pulsating flow between parallel surfaces.

The second category of exact solutions is that for which the nonlinear convective term is not identically zero. Examples of these types of flows include stagnation-point flow, the flow in convergent and divergent channels, and the flow over a porous wall.

## 4.2 Low-Reynolds-number solutions to the Navier-Stokes equation – Stokes equations

Since the *Reynolds number* ($Re = \frac{\rho U L}{\mu}$, where $L$ is the characteristic length, $U$ is a characteristic velocity, and the other quantities have been already defined) may be interpreted as the ratio of the inertia forces of the fluid to the viscous forces, flows at very small Reynolds numbers amounts to the condition of negligible inertia forces. This implies that the convective inertia term $(\boldsymbol{u} \cdot \boldsymbol{\nabla})\boldsymbol{u}$ in the N-S equation is assumed to be small compared with the viscous term, resulting in the *Stokes equations*.

A famous result obtained from the Stokes equations is the famous *Stokes' drag law*. Here it is found that there is a connection between force and viscosity for a free-falling sphere in a viscous fluid that is used in the falling-sphere viscometer, which is a device that measures the viscosity of a liquid by measuring the time required for a spherical ball to fall a certain distance under gravity through a tube filled with the fluid whose viscosity is to be determined.

## 4.3 The laminar boundary layer

The boundary layer is a thin region close to the surface of a body where viscous effects are significant. The boundary layer is formed in internal flows, such as that at the inlet of a tube, and in external flows, such as the flow over an aircraft wing.

In the case of an airfoil, the boundary layer at first flows smoothly over the streamlined shape of the airfoil, and the flow in the boundary layer is laminar. As the flow approaches the center of the wing, it begins to lose speed due to friction and the boundary layer becomes thicker and turbulent.

No solution does exist for the laminar boundary layer on airfoil of arbitrary shape. Nonetheless, it is possible to obtain a solution from a simplified form of the N-S equation for a flat plate subjected to a uniform flow—indeed, an oversimplification to a complex problem.

5. **Reynolds averaged Navier-Stokes equations**

As we just saw, the N-S equation affords no solution to the more complex laminar flows, and as mentioned earlier, it is not applicable in its current form to turbulent flows, where lies most flows of interest.

Similar to viscous stresses, the turbulent velocity fluctuations also produce stresses, the so-called *Reynolds stresses*, which are the mean forces (per unit area) imposed on the mean flow by turbulent fluctuations.

The Reynolds-averaged Navier–Stokes equations (RANS equations) are an extension of the regular N-S equation, applicable to turbulent flows. They are obtained by means of the so-called *Reynolds decomposition*, whereby the instantaneous velocities are decomposed into their time-averaged and fluctuating quantities.

The RANS equations have the same structure of the laminar N-S equation, with the difference that the velocities in the RANS equations are represented by time-averaged values. However, an additional term appears in the RANS equations, which represents the effect of turbulence. This term introduces six unknown terms associated with the velocities fluctuations, but the matching equations are not available to close the system. Therefore, this term needs to be modelled in order to close the system of equations. Several approaches have evolved for this purpose.

Computational Fluid Dynamics (CFD) is a branch of fluid mechanics that applies numerical analysis and data structures to analyse and solve problems that involve turbulence. Computers programs are used to perform the calculations required to simulate the free-stream flow and the interaction of the fluid (liquids and gases) with surfaces defined by boundary conditions. Thanks to high-speed computers, solutions can be achieved for the largest and most complex problems.

CFD applies numerical simulation of turbulence. This can be essentially done by Direct Numerical Simulation (DNS) and by Large Eddy Simulation (LES). DNS explicitly resolves and captures all scales of turbulence, including the smallest ones. The principal idea behind LES is to reduce the computational cost by ignoring the smallest length scales, which are the most computationally expensive to resolve. For most engineering applications, it is unnecessary to resolve the details of the turbulent fluctuations, and, therefore, the RANS simulation is preferable.

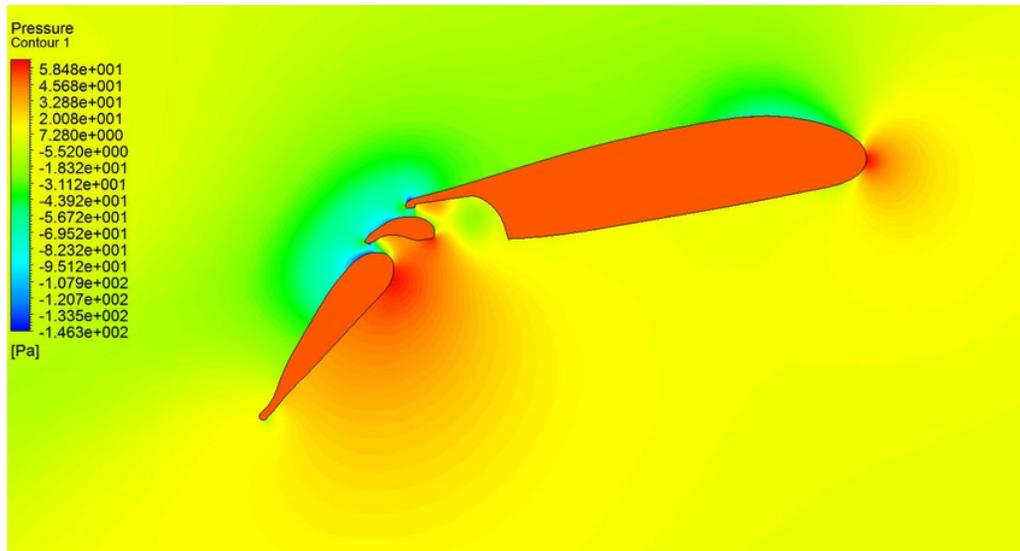

Pressure distribution around an aircraft airfoil obtained by a CFD simulation[1]

The application of the RANS equations requires the adoption of a *turbulence model*, which is a computational procedure to close the system of mean flow RANS equations. Turbulence models allow the calculation of the mean flow without first calculating the full time-dependent flow field. A turbulence model provides expressions for the Reynolds stresses, and from an engineering point of view, it must be simple, accurate and economical to run.

**6. Impact of the Navier-Stokes equation on the evolution of Fluid Dynamics**

It soon became clear right after the first appearance of the N-S equation that for this equation to become more useful, it had to address the phenomenon of turbulence. It was Osborne Reynolds, from whom the popular *Reynolds number* borrows the name, who in 1895 proposed the inclusion of turbulence by averaging of the N-S equation over the turbulent fluctuations, which resulted in the Reynolds-averaged Navier–Stokes equations, discussed earlier.

The next major advance was turbulence modelling, which was initially done by *ad-hoc* procedures. Such was the case of the *mixing length* concept introduced by Prandtl in 1926. This is the simplest turbulence model, and still used in RANS simulations by CFD.

In 1933 Nikuradse [6] published "Laws of flow in rough pipes", which was based on his carefully measurements of friction that a turbulent flow experiences as it flows through a rough pipe. These measurements were interpolated by Colebrook which resulted in the formula that bears his name and is the accepted design formula for turbulent friction. Its plot in 1944 by Moody resulted in the well-known *Moody chart* for pipe friction, which is considered the most famous useful figure in Fluid Dynamics.

As far as applications are concerned, the N-S equation in its original form certainly has limitations in its applicability, nonetheless, since its first appearance it has fostered an enormous amount of work to understand, to model and to predict fluid behavior, which has decisively contributed to the evolution of Fluid Dynamics.

---

[1] Used with the permission of Ahmed Al Makky, consultant for the CFD industry, at his website: 'Computational Fluid Dynamics is the Future', at https://cfd2012.com/wings.html

## 7. Conclusions

The 200 years of the Navier-Stokes equation's existence have left an indelible mark on the field of Fluid Dynamics. From its inception in the early 19th century to its continued relevance in the 21st century, this equation has been instrumental in expanding our understanding of fluid motion. The contributions of Claude-Louis Navier, George Stokes, and subsequent scientists have solidified its place as a fundamental pillar in science and engineering.

The equation's evolution from its original form to the Reynolds-averaged Navier-Stokes equations and beyond has allowed us to tackle a wide range of fluid flow problems, from laminar to turbulent flows. Its application in Computational Fluid Dynamics (CFD) has opened doors to simulating complex fluid behaviors, aiding in the design of aircraft, vehicles, power stations, and various fluid flow machines.

While the Navier-Stokes equation has its limitations, particularly in modeling turbulent flows, it has been a catalyst for research and innovation. The quest to understand and predict fluid behavior has led to the development of turbulence models, numerical methods, and experimental techniques. This ongoing pursuit continues to shape the future of Fluid Dynamics.

As we celebrate the bicentennial of this equation, we acknowledge its enduring significance and the countless contributions it has inspired. From fundamental research to practical applications, the Navier-Stokes equation remains a testament to the enduring power of scientific inquiry and its impact on our understanding of the natural world.

**Note**

The present work was adapted from a contribution of the present author to parts of the brochure COMPUTATIONAL FLUID DYNAMICS - A Historic Mechanical Engineering Landmark to be published by The American Society of Mechanical Engineers—ASME.

**Biographical Sketches**

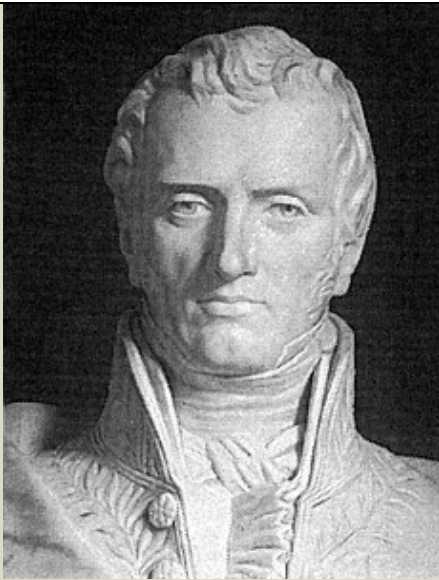

Bust of Claude-Louis Marie Henri Navier at the *École des Ponts et Chaussées*, Paris
(10 February 1785–21 August 1836).

Claude-Louis Navier was an engineer trained first at the *École Polytechnique* and then at the *École des Ponts et Chaussées*. He embodied a new style of engineering that combined the analytical skills acquired at the *Polytechnique* with the practical bent of the *Écoles d'application*. Through his theoretical research and his teaching he contributed to a renewal of the science of mechanics that forged a much better fit between it and the needs of engineers.

In 1824, Navier was admitted into the French Academy of Science. In 1830, he took up a professorship at the *École Nationale des Ponts et Chaussées*, and in the following year he succeeded Cauchy as professor of calculus and mechanics at *Polytechnique (Adapted from the Wikipedia https://en.wikipedia.org/wiki/Claude-Louis_Navier, Accessed Dec. 11, 2021)*

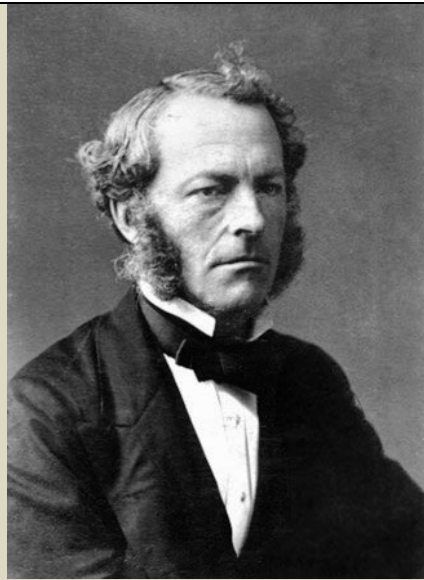

Sir George Gabriel Stokes-circa 1860
Pembroke College, Cambridge
(13 August 1819–1 February 1903).

George Stokes spent his entire career at the University of Cambridge, where he was the Lucasian Professor of Mathematics from 1849 until his death in 1903. As a physicist, Stokes made seminal contributions to Fluid Mechanics, including the Navier–Stokes equations.

Stokes was made a baronet (hereditary knight) by the British monarch in 1889. In 1893 he received the Royal Society's Copley Medal, then the most prestigious scientific prize in the world.

He represented Cambridge University in the British House of Commons from 1887 to 1892, sitting as a Conservative. Stokes also served as president of the Royal Society from 1885 to 1890 and was briefly the Master of Pembroke College, Cambridge. *(Adapted from the Wikipedia https://en.wikipedia.org/wiki/Sir_George_Stokes,_1st_Baronet, Accessed Dec. 11, 2021)*